\documentclass[twocolumn]{autart}    

\usepackage{graphicx}          
\usepackage{amsmath}
\usepackage{mathtools}
\usepackage{hyperref}
\usepackage{multirow}
\usepackage{xcolor}

\edef\endfrontmatter{%
  \unexpanded\expandafter{\endfrontmatter}
  \noexpand\endNoHyper 
}

\newcommand{\norm}[1]{|| #1 ||}
\newcommand{\Hone}{\mathcal{H}_{1}}
\newcommand{\Htwo}{\mathcal{H}_{2}}
\newcommand{\Hinf}{\mathcal{H}_{\infty}}

\begin{document}

\begin{frontmatter}

\title{Data-driven estimation of system norms via impulse response} 


\author[UDESC]{Luan Vinícius Fiorio}\ead{luan.lvf@edu.udesc.br},    
\author[UFRGS]{Chrystian Lenon Remes}\ead{chrystian.remes@ufrgs.br},  
\author[UFRGS]{Lucíola Campestrini}\ead{luciola@ufrgs.br},
\author[UDESC]{Yales Rômulo de Novaes}\ead{yales.novaes@udesc.br}  

\address[UDESC]{Santa Catarina State University, R. Paulo Malschitzki, 200 - Zona Industrial Norte, Joinville SC, 89219-710, Brazil}  
\address[UFRGS]{Federal University of Rio Grande do Sul, Av. Osvaldo Aranha, 103 - Centro Histórico, Porto Alegre RS, 90035-190, Brazil}             


\begin{abstract}                          
This paper proposes a method for estimating the norms of a system in a pure data-driven fashion based on their identified Impulse Response (IR) coefficients using only a single batch of data (one-shot). The calculation of norms is briefly reviewed and the main expressions for the IR-based estimations are presented. As a case study, the $\Hone$, $\Htwo$, and $\Hinf$ norms of the sensitivity transfer function of five different discrete-time closed-loop systems are estimated for a Signal-to-Noise-Ratio (SNR) of 10 dB, achieving low percent error values if compared to the real value. To verify the influence of the noise amplitude, norms are estimated considering a wide range of SNR values, for a specific system, presenting low Mean Percent Error (MPE) if compared to the real norms. The proposed technique is also compared to an existing state-space-based method in terms of $\Hinf$, through Monte Carlo experiments, showing a reduction of approximately 48 \% in the MPE for a wide range of SNR values.
\end{abstract}

\end{frontmatter}

\section{Introduction}

A system norm is a single number that contains information regarding gain, energy, robustness, among other possible physical interpretations, which can be used as a tool for system analysis and controller design \cite{Skogestad&Postlethwaite:2005}. In a case where the model of a process is not available or is too complex to be obtained, an alternative data-driven method can be used to estimate system norms.

The most common data-driven approach is to obtain the norms via frequency response \cite{Tacx:2021}, but it usually requires more than one experiment to acquire the necessary data. Iterative methods as Power Iterations and Weighted Thompson Sampling \cite{Muller:2020} are also a viable solution at the expense of high computational cost. A time domain approach for $\Hinf$ estimation with the Toeplitz matrix of the system's Markov parameters is presented in \cite{Oomen:2014}, and a different computation to the same approach is shown in \cite{GoncalvesDaSilva:DD-Certification-2020}, based on the system's subspace (state-space) identification.

This work, inspired by \cite{GoncalvesDaSilva:DD-Certification-2020}, presents a time domain data-driven and one-shot method for estimating the norms of a Single-Input Single-Output (SISO) system solely relying on its estimated Impulse Response (IR) coefficients. In order to achieve a lower estimation error than state-space-based methods with corrupted data, the IR is identified through regularized least squares \cite{ChenLjung:2012}. Solutions are presented for norms $\Hone$, $\Htwo$, and $\Hinf$.
\section{Norm estimation by impulse response}
\label{sec:norms}
The output signal $y(k)$ of a linear discrete-time causal and SISO system $G$ is given by
\begin{equation}
		G : y(k) = g(k)\ast u(k) = \sum_{n=0}^{\infty}g(k-n)u(n),
		\label{eqA:gk}
\end{equation}
which is the convolution between the input signal $u(k)$ and the signal that represents the impulse response of the system, $g(k)$. In other words, knowing $g(k)$ is enough to characterize any input-output relation of a linear system $G$. Other important definitions are the norms of signals and systems, widely used to characterize their measurements. According to \cite[A.5]{Skogestad&Postlethwaite:2005}, the norms $\mathcal{L}_p$, $\mathcal{L}_\infty$, used for signals, can be defined as

\begin{equation}
\label{eq:lp}
	\mathcal{L}_p: \norm{x(k)}_{p} = \left(\sum_{k=0}^{\infty}|x(k)|^p \right)^{1/p},
\end{equation}
\begin{equation}
\label{eq:l2}
	\mathcal{L}_\infty: \norm{x(k)}_{\infty} = \max |x(k)|,
\end{equation}
where $x(k)$ is any time domain signal, and $p$ can assume different values, typically 1 or 2. The $\Hone$, $\Htwo$ and $\Hinf$ norms used for SISO systems, also from \cite[A.5]{Skogestad&Postlethwaite:2005}, are defined as
\begin{equation}
\label{eq:g1}
	\Hone : \norm{G}_1 = \sum_{k=0}^{\infty}|g(k)| = \max_{u(k)\neq 0}\dfrac{\norm{g(k)\ast u(k)}_{\infty}}{\norm{u(k)}_{\infty}},
\end{equation} 
\begin{equation}
\label{eq:g2}
	\Htwo : \norm{G}_2 = \sqrt{\sum_{k=0}^{\infty}|g(k)|^2} = \norm{g(k)}_{2},
\end{equation}
\begin{equation}
\label{eq:ginf}
	\Hinf : \norm{G}_{\infty} = \max_{u(k)\neq 0}\dfrac{\norm{g(k)\ast u(k)}_{2}}{\norm{u(k)}_{2}}.
\end{equation}
In the case of a stable system, it is known that $\lim_{k\to\infty}g(k) = 0$. In order to use limited data for estimating system norms, the IR terms with an order greater than $M$ are assumed to be negligible. Therefore, the convolution in \eqref{eqA:gk} can be truncated at $M$ terms, resulting in the approximation
\begin{equation}
		G : y(k) = \sum_{n=0}^{\infty}g(k-n)u(n) \approx \underbrace{\sum_{n=0}^{M}g(k-n)u(n)}_{|g(M+1)|< \epsilon\text{, with }\epsilon\to 0^+}.
		\label{eqA:gk_approx}
\end{equation}
Notice that only the $M$ first elements of the IR are considered to be sufficient to characterize the input-output relation of a system. In terms of $\Hone$ and $\Htwo$ norms, from \eqref{eq:g1} and \eqref{eq:g2}, considering the truncated convolution in \eqref{eqA:gk_approx}, the following approximations can be obtained:
\begin{equation}
\label{eq:norm_G1}
	\norm{G}_1 \approx \sum_{k=0}^{M}|g(k)|,
\end{equation}
\begin{equation}
\label{eq:norm_G2}
    \norm{G}_2 \approx \sqrt{\sum_{k=0}^{M}|g(k)|^2}.
\end{equation}

For the $\Hinf$ norm, which represents the maximum relation of gain considering all sets of input signals, expression \eqref{eq:ginf} cannot be directly used, since it requires to consider all possible input signals in its evaluation. An alternative strategy is to obtain a matrix relation for $g(k)$, allowing for the use of induced norm properties. Expanding the relation given in \eqref{eqA:gk_approx} to the $M$ first terms,
\begin{equation}
	\begin{cases}
		& y(0) = g(0) u(0)\\
		& y(1) = g(1) u(0) + g(0) u(1)\\
		& \cdots \\
		& y(M) = g(M)u(0) + \cdots + g(0) u(M),
	\end{cases}
	\label{eqA:G}
\end{equation}
the following matrix relation is obtained:
\begin{equation}
	\underbrace{
	\begin{bmatrix}
		y(0) \\ y(1) \\ \cdots \\ y(M)
	\end{bmatrix}
	}_{Y_M}
	 = 
	\underbrace{
	\begin{bmatrix}
		g(0) 		& 0 			& \cdots &		0		\\
		g(1) 		& g(0)		&	\cdots &		0		\\
		\vdots 	& \vdots	& \ddots & \vdots \\
		g(M)		& g(M-1)	& \cdots & 	g(0)
	\end{bmatrix}
	}_{G_M}
	\underbrace{
	\begin{bmatrix}
		u(0) \\ u(1) \\ \cdots \\ u(M)
	\end{bmatrix} 
	}_{U_M}
	.
	\label{eqA:GM}
\end{equation}
It should be noted that, in \eqref{eqA:GM}, matrix $G_M$ represents the terms of the system's impulse response, in the form of a Toeplitz matrix, and that its multiplication with the input vector $U_M$, truncated at $M$ elements, results in the output vector $Y_M$. From the assumption that $M$ is sufficiently high, it can be said that matrix $G_M$ characterizes the IR $g(k)$ and, consequently, system $G$.

The characterization of system $G$ in a matrix form, represented by $G_M$, allows the use of some matrix properties, as it is the case of the induced norm, \cite[A.5]{Skogestad&Postlethwaite:2005}:
\begin{equation}
		\norm{G_M}_{ip} = \max_{U_M\neq 0}\dfrac{\norm{G_MU_M}_{p}}{\norm{U_M}_{p}},
	\label{eqA:ind_norm}
\end{equation}
where the subscript $i$ stands for induced. In short, $\norm{G_M}_{ip}$ is a form of representing the gain of the system $G$ considering a set of possible input signals $U_M$. Among the possible induced norms, the induced-2 norm can be highlighted:
%
\begin{equation}
\label{eq:gmi2}
    \norm{G_M}_{i2} = \bar{\sigma}(G_M) = \sqrt{\lambda_{max}\left(G_M' G_M\right)},
\end{equation}
where $\bar{\sigma}$ and $\lambda_{max}$ stands for largest singular value and largest eigenvalue, respectively. Note that \eqref{eq:gmi2} is evaluated directly with $G_M$, which can be obtained with a single input signal.
%
Comparing \eqref{eq:gmi2} to the $\Hinf$ norm definition from \eqref{eq:ginf}, knowing that convolution \eqref{eqA:gk_approx} is equivalent to the matrix expression in \eqref{eqA:GM}, it is observed that
\begin{equation}
    \norm{G}_\infty \approx \max_{U_M \neq 0}\frac{\norm{G_M U_M}_{2}}{\norm{U_M}_{2}} = \norm{G_M}_{i2}.
\end{equation}
%
Consequently, the calculation of the $\Hinf$ norm can be approximated by
%
\begin{equation}
    \norm{G}_\infty \approx \bar{\sigma}(G_M) = \sqrt{\lambda_{\max}(G_M' G_M)}.
    \label{eq:norm_Ginf}
\end{equation}

Note that in \eqref{eq:norm_G1}, \eqref{eq:norm_G2}, and \eqref{eq:norm_Ginf}, only the knowledge of the $M$ first elements of the impulse response is sufficient to estimate the system norms. Using this idea, one-shot algorithms for the IR estimation can be used to estimate the coefficients of $g(k)$, and consequently, the norms of $G$. An option for a more precise estimation of the IR is to use regularization, for the following reasons: i) the variance of the estimates increases proportionally with the order $M$, forcing the user to compensate this fact by reducing the number of estimated coefficients, problem that is suppressed with the use of regularization \cite{ChenLjung:2012}; and ii) regularization is well-known to improve estimates of sparse signals \cite{Brunton:2019}, and the IR is a sparse signal for sufficiently high $M$. The regularized least squares for impulse response estimation is described in \cite{ChenLjung:2012,Chen&Ljung:Implementation-Regularization-2013}, and it is available in MATLAB\textregistered~\cite{Matlab:2017}, Python~\cite{Fiorio:2021}, and R~\cite{Yerramilli:2017}, presenting a viable solution for such problem.

\section{Case study}
\label{sec:case_study}

The objective of this case study is to obtain $\Hone$, $\Htwo$ and $\Hinf$ norms of the sensitivity transfer function $S(z)$ of a closed-loop discrete-time system, where $z$ is the discrete time-shift operator. In practice, the norm $\norm{S(z)}_{\infty}$ is used as a measure of robustness \cite{Skogestad&Postlethwaite:2005}. In a data-driven case, for example, $S(z)$ cannot be obtained with models since they are not available. Assuming an experiment where a given reference signal $r(k)$ is applied to the closed-loop system, generating an output $y(k)$, being
\begin{equation}
    y(k) = T(z) r(k) + S(z)n(k),
\end{equation}
where
\begin{equation}
    T(z) = \frac{C(z)G(z)}{1 + C(z)G(z)}, \quad S(z) + T(z) = 1,
\end{equation}
and $n(k)$ is the output noise. It follows that $y(k) = [1 - S(z)]r(k) + S(z)n(k) = r(k) + S(z)[n(k) - r(k)]$. Consequently $y(k) - r(k) = S(z)[n(k) - r(k)]$, resulting in $r(k) - y(k) = C^{-1}(z) u(k) = S(z) [r(k) - n(k)]$. Henceforth, the input-output set $\{r(k),C^{-1}(z)u(k),k=1...N \}$ can be acquired from the process and used for the estimation of the impulse response of $S(z)$, $s(k)$. Note that $n(k)$ is not considered in the data set since it cannot be acquired. The norms $\norm{S(z)}_{1}$ and $\norm{S(z)}_{2}$ can be directly obtained by expressions \eqref{eq:norm_G1} and \eqref{eq:norm_G2}, respectively. In order to estimate the $\Hinf$ norm of $S(z)$, a Toeplitz matrix $S_M$ can be built with its IR, as in \eqref{eqA:GM}. Finally, from \eqref{eq:norm_Ginf}, $\norm{S(z)}_{\infty} \approx \norm{S_M}_{i2}$, with $M$ sufficiently large.
\subsection{Examples}


To illustrate the proposed method, five closed-loop systems - plant $G(z)$ and controller $C(z)$ - are used as subjects, since they represent the structure of systems that are commonly found in engineering problems, which are presented in Table~\ref{tab:systems}. Table~\ref{tab:super_table} shows the $\Hone$, $\Htwo$ and $\Hinf$ norms of $S(z)$ calculated by model, assumed here as the real values, and estimated by the proposed data-driven method. The reference signal is a Pseudo-Random Binary Signal (PRBS) with $N=2000$ samples, and the length of the IR was arbitrarily set as $M=100$. The IR is estimated in a regularized fashion with the Tuned-Correlated (TC) kernel \cite{Chen&Ljung:Implementation-Regularization-2013}. Additive white Gaussian noise with zero mean and a Signal-to-Noise-Ratio (SNR) of 10 dB was included in the process output $y(k)$ and fed back to the system. The percent error between the real and estimated value is also shown in Table~\ref{tab:super_table} and indicates good performance of the proposed technique for all cases, since the obtained error values are low.

\begin{figure} [!ht]
	\centering
	\caption{Estimated $\Hone$, $\Htwo$ and $\Hinf$ norms of $S(z)$, for System 2, with its real values in dashed lines, for a wide SNR range.}
	\includegraphics[width=0.94\linewidth]{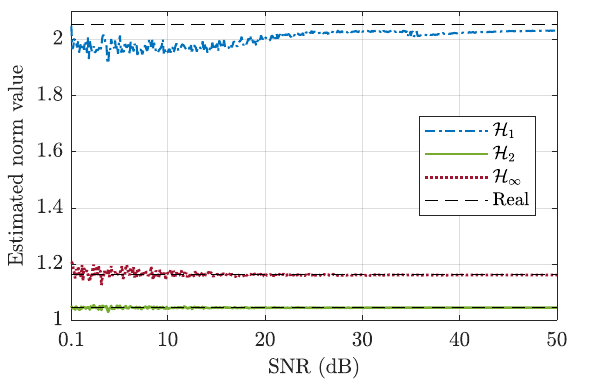} \\
	\label{fig:norms_by_snr}
\end{figure}

Now, in order to verify the influence of the noise amplitude, $\Hone$, $\Htwo$, and $\Hinf$ norms of $S(z)$ for System 2 are estimated with an SNR value varying from 0.1 to 50~dB with a step of 0.1~dB for the noise at the output, as it is shown in Figure~\ref{fig:norms_by_snr}. The estimations are done using the same $r(k)$ as aforementioned. The Mean Percent Error (MPE) between estimated and real norm is used as a measure of precision, given as the average of the errors found for each experiment for all SNR values. The estimation of $\Hone$ for the whole tested SNR range presents an MPE of 2.4287~\%, whilst for the norm $\Htwo$ the MPE is of 0.0903~\%, and for $\Hinf$ the MPE achieves 0.3729~\%.
%

\begin{figure} [!ht]
	\centering
	\caption{Mean value of 100 Monte Carlo runs for the proposed and a state-space method \cite{GoncalvesDaSilva:DD-Certification-2020}, for estimating $\norm{S(z)}_{\infty}$ of System 1, as well as its box plot representation.}
	\includegraphics[width=0.94\linewidth]{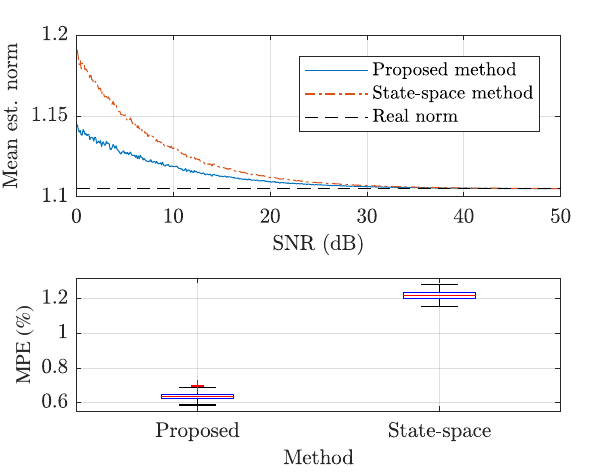} \\
	\label{fig:hinf_comp}
\end{figure}

For the sake of comparison, Figure~\ref{fig:hinf_comp} presents a Monte Carlo experiment of 100 different noise realizations for each SNR value of the estimated $\Hinf$ norm of $S(z)$ for System~1, considering $M=100$ for all cases, by the technique proposed in this paper in a regularized fashion, and by the state-space-based technique proposed in \cite{GoncalvesDaSilva:DD-Certification-2020}. The approach proposed in this work obtained an MPE for the Monte Carlo experiment of 0.6351~\%, whilst the state-space approach achieved 1.2187~\% of MPE, showing a reduction of 47.8871~\% in MPE.
\section{Conclusion}
\label{sec:conclusion}

This paper presented a one-shot method for estimating the $\Hone$, $\Htwo$, and $\Hinf$ norms of a system, using time domain signals, relying solely on its estimated impulse response coefficients. A closed loop case study with the objective of estimating the $\Hone$, $\Htwo$, and $\Hinf$ norms of $S(z)$ is presented for five different systems. All data-driven estimated norms were close to the real norms, showing a maximum error of 5.2485 \% - for System 2, norm $\Hone$, and $M = 100$. For a wide range of SNR, norms were estimated for System 2, presenting a low MPE for all cases. In a Monte Carlo comparison to a state-space-based technique, for the $\Hinf$ norm, the proposed method has shown to reduce the MPE by 47.8871~\%. As for future related research subjects, are suggested: finding bounds for the estimation error; automatic choice of the estimated IR length $M$.
\begin{ack}                               
This study was financed in part by the Coordenação de Aperfeiçoamento de Pessoal de Nível Superior - Brasil (CAPES) - Finance Code 001 and partly by the Fundação de Amparo à Pesquisa e Inovação do Estado de Santa Catarina (FAPESC) - Grant number 288/2021.  
\end{ack}
\begin{table*}[t]
  \caption{System's transfer functions $G(z)$ and controllers $C(z)$ used as examples.}
  \centering
  \begin{tabular}{c c c}
    \hline
    \textbf{System} & $\mathbf{G(z)}$ & $\mathbf{C(z)}$ \\
    \hline
    1 & $\frac{0.5}{(z - 0.9)}$ & $\frac{0.3797 (z-0.9)}{(z-1)}$ \\
    2 & $\frac{-0.1 (z - 0.5)}{(z - 0.9)(z - 0.8)}$ & $\frac{-1.1600 (z-0.9719)}{(z-1)}$ \\    
    3 & $\frac{-0.05 (z - 0.6)}{(z^2 - 1.8z + 0.82)}$ & $\frac{-3.7144 (z-0.9351) (z-0.4210)}{z(z-1)}$ \\
    4 & $\frac{-0.05 (z - 1.4)}{(z - 0.9)(z - 0.8)}$ & $\frac{4.7942 (z-0.9) (z-0.8)}{z(z-1)}$ \\
    5 & $\frac{3.605 (z-0.55)(z^2 - 1.62 z + 0.6586)}{(z^2 - 1.84z + 0.8564)(z^2 - 1.26z + 0.4069)}$ & $\frac{0.0519 (z-0.8977)}{(z-1)}$ \\
    \hline    
  \end{tabular}
  \label{tab:systems}
\end{table*}
\begin{table*}[t]
	\caption{Each system's $\Hone$, $\Htwo$, and $\Hinf$ norm of $S(z)$ calculated by model (Real), estimated via data (Data), and its percent error.}
	\centering
	\begin{tabular}{c @{\hskip 10mm} c c c @{\hskip 10mm} c c c @{\hskip 10mm} c c c}
	      & & $\Hone$ & & & $\Htwo$ & & & $\Hinf$ & \\
		\hline
    		\textbf{System} & \textbf{Real} & \textbf{Data} & \textbf{Error (\%)} & \textbf{Real} & \textbf{Data} & \textbf{Error (\%)} & \textbf{Real} & \textbf{Data} & \textbf{Error (\%)}  \\
		\hline
		1 & 2.0000 & 2.0076 & 0.3825 & 1.0511 & 1.0520 & 0.0784 & 1.1049 & 1.1210 & 1.4635\\
		2 & 2.0544 & 1.9466 & 5.2485 & 1.0441 & 1.0428 & 0.1293 & 1.1619 & 1.1521 & 0.8408\\
		3 & 2.1656 & 2.1496 & 0.7358 & 1.0542 & 1.0509 & 0.3166 & 1.1272 & 1.1331 & 0.5186\\
		4 & 2.4794 & 2.4778 & 0.0649 & 1.0811 & 1.0846 & 0.3175 & 1.5348 & 1.5375 & 0.1767\\
		5 & 2.0307 & 1.9970 & 1.6578 & 1.0523 & 1.0543 & 0.1889 & 1.1006 & 1.1224 & 1.9781\\
		\hline
	\end{tabular}
	\label{tab:super_table} 
\end{table*}
\bibliographystyle{unsrt}        
\bibliography{root}           



\appendix
\end{document}